%% file: BayesianRegistration.tex
\documentclass[aoas]{imsart}

\RequirePackage{amsthm,amsmath,amsfonts,amssymb}
\RequirePackage[authoryear]{natbib}
\RequirePackage[colorlinks,citecolor=blue,urlcolor=blue]{hyperref}
\RequirePackage{graphicx}

\usepackage[linesnumbered,ruled,vlined]{algorithm2e}

\startlocaldefs
\newcommand{\one}{\mathbf{1}}

\newcommand{\C}{\mathbf{C}}
\newcommand{\I}{\mathbf{I}}
\newcommand{\K}{\mathbf{K}}
\newcommand{\R}{\mathbf{R}}
\newcommand{\Y}{\mathbf{Y}}
\newcommand{\Sbf}{\mathbf{S}}

\newcommand{\e}{\mathbf{e}}
\newcommand{\w}{\mathbf{w}}
\newcommand{\bR}{\mathbf{R}}
\newcommand{\bY}{\mathbf{Y}}
\newcommand{\bw}{\mathbf{w}}

\newcommand{\bfp}{\mathbf{p}}

\newcommand{\Reg}{\text{Reg}}
\newcommand{\A}{\mathbf{A}}
\newcommand{\s}{\mathbf{s}}
\newcommand{\vect}[1]{\text{vec}(#1)}

\endlocaldefs

\newcommand{\beginsupplement}{%
        \setcounter{table}{0}
        \renewcommand{\thetable}{S\arabic{table}}%
        \setcounter{figure}{0}
        \renewcommand{\thefigure}{S\arabic{figure}}%
     }

\begin{document}
\begin{frontmatter}
\title{Bayesian Functional Registration of fMRI Activation Maps}

\begin{aug}

\author{\fnms{Guoqing} \snm{Wang}},
\author{\fnms{Abhirup} \snm{Datta}}
\and
\author{\fnms{Martin A.}  \snm{Lindquist}\ead[label=e1]{mlindqui@jhu.edu}}
\address{Department of Biostatistics,
Johns Hopkins University,
\printead{e1}}

\end{aug}



\begin{abstract}
Functional magnetic resonance imaging (fMRI) has provided invaluable insight into our understanding of human behavior. However, large inter-individual differences in both brain anatomy and functional localization \emph{after} anatomical alignment remain a major limitation in conducting group analyses and performing population level inference.  This paper addresses this problem by developing and validating a new computational technique for reducing misalignment across individuals in functional brain systems by spatially transforming each subjects functional data to a common reference map. Our proposed Bayesian functional registration approach allows us to assess differences in brain function across subjects and individual differences in activation topology.  It combines intensity-based and feature-based information into an integrated framework, and allows inference to be performed on the transformation via the posterior samples. We evaluate the method in a simulation study and apply it to data from a study of thermal pain. We find that the proposed approach provides increased sensitivity for group-level inference. 
\end{abstract}


\end{frontmatter}

\section{Introduction}

In recent years functional magnetic resonance imaging (fMRI) has provided invaluable insight into understanding the neurophysiological underpinnings of human behavior. Much of these gains have been driven by the combination of increasingly sophisticated statistical and computational techniques and the ability to capture brain data at finer spatial and temporal resolution.
Together, these advances are allowing researchers to study brain-behavior correlations and develop population-level models of the functional brain representations underlying behavior, performance, clinical status and prognosis. However, a fundamental obstacle to using emerging analytic techniques effectively in these settings is individual variation in functional brain anatomy. That is, the fact that the precise locations of brain activation are not perfectly aligned across subjects. Addressing this obstacle will greatly extend the capability to perform group-level hypothesis testing, use machine learning to develop population-level brain models that predict behavior, and take advantage of the full spatial resolution of fMRI.

In a task-based fMRI study, each subject is typically administered one or more stimuli and observed at hundreds of time points. At each time point, the subject’s blood oxygenation level dependent (BOLD) response is recorded at roughly 100,000 spatial locations (voxels), giving rise to multivariate time series data. In order to properly perform statistical analysis over multiple subjects, it is necessary for each voxel to lie within the same brain structure for each subject. This does not occur naturally, as all subjects have brains of different sizes and shapes. Therefore, to achieve this goal all subjects’ brains are normalized to a stereotaxic space prior to analysis. 

Typically, normalization is performed using a high-resolution anatomical scan that is acquired in the same session and spatially aligned with the fMRI data. Current standard practice is to use nonlinear transformations to warp individual subjects’ anatomical scan to an average, anatomically based 3D reference space (e.g., the “Montreal Neurologic Institute” (MNI) space). These transformations are then applied to the fMRI data to ensure they lie in MNI space, and data can be compared across subjects  \citep{lindquist2008statistical, ombao2016handbook}.

While this procedure efficiently makes the brains of all subjects overlap with regards to gross anatomical landmarks, it does not implicitly take into consideration residual differences in brain anatomy, and the location and distribution of functional regions relative to these landmarks.  For example, primary visual cortex can vary in size by as much as 2-fold across different subjects' brains \citep{rademacher1993topographical} and in location relative to other anatomical landmarks \citep{amunts2000brodmann}, as can sulcal locations \citep{thompson1996three}. In some cases, inter-individual differences can be accounted for by normalization, but in other cases they cannot. For example, the cingulate sulcus is structurally dimorphic \citep{vogt1995human} —a substantial minority of individuals possesses a double cingulate sulcus—precluding anatomical alignment using current methods. In addition, considerable variability in functional localization persists even after anatomical alignment \citep{duncan2009consistency}. For example, the location of visual motion-sensitive area MT can vary by more than $2 \, cm$. The location and extent of lateral occipital cortex \citep{iordan2016locally} and fusiform face area \citep{allison1999electrophysiological, mccarthy1999electrophysiological} are also highly variable across individuals and located in different places relative to anatomical sulcal landmarks. 

Large inter-individual differences in both brain anatomy and functional localization after anatomical alignment are a major limitation in conducting group analyses and population level inference. This is true because a large number of statistical models are performed voxel-wise (i.e., subjects are compared using a separate model at each voxel). This can lead to a situation where voxels from different functional brain regions are compared to one another in a model, significantly complicating interpretation. In addition, voxels are often used as features in prediction models, and any variation in the location of functional regions across subjects would lead to feature misalignment. Finally, it degrades the quality of the data.  Though individual-subject, whole-brain fMRI data can now be collected at a resolution of $2 \, mm^3$ voxels or smaller, once functional maps are combined across individuals to construct a population-level model, the effective spatial resolution is dramatically lower.

Recent approaches have sought to circumvent this problem by mapping individual brains into a functional population-level reference space rather than an anatomically based brain space. The first such model is the ‘hyperalignment’ procedure \citep{Haxby2011}. Here brain activity patterns corresponding to stimuli and other cognitive events are represented as vectors in a neural representational space spanned by the voxels in a local neighborhood. Hyperalignment rotates each participant’s local voxel-wise activity patterns through multivariate voxel space using a Procrustes transformation to align the representational geometry across subjects. Mathematically, this is similar to Canonical Correlation Analysis (CCA). A number of refinements of hyperalignment have recently appeared, including: kernel hyperalignment \citep{Lorbert2012}, which performs nonlinear hyperalignment in an embedding space; regularized hyperalignment \citep{Xu2012} which uses a ridge CCA formulation; the two-phase joint SVD-Hyperalignment algorithm \citep{Chen2014}, where singular value decomposition (SVD) provides a lower dimensional feature space where hyperalignment is applied; the shared response model \citep{Chen2014} which casts the model in a probabilistic framework; and searchlight hyperalignment \citep{Guntupalli2016} which allows for whole-brain coverage. Other methods have been developed that align subjects based on inter-subject correlations in time-series data during movie viewing. One method is functional time series alignment \citep{Sabuncu2010}, which matches voxels across subjects using a 2D ‘rubber sheet’ warping and maximizing inter-subject correlations across voxels \citep{Hasson2004}. Another is functional connectivity alignment \citep{Conroy2013}, which matches voxels to a functional reference by minimizing the Frobenius norm of the difference between a subject’s connectivity matrix and a reference matrix with a shape-preserving penalty function. Both methods use warping and penalization strategies specific to the cortex, so it is unclear how easily they can be adapted to include subcortical regions. 

This body of work shows great promise for allowing participants to vary in their functional activation patterns \citep{Haxby2011}, but suffers from several shortcomings. First, hyperalignment requires subjects to watch a long film (up to 2 hours) to align subjects into a common representational space, and functional connectivity-based methods require substantial resting state data. Second, the choice of reference data influences which types of functional patterns can be appropriately aligned. For example, movie reference data works well for audio-visual representations but not prefrontal and limbic networks. Functional connectivity-based measures work better for limbic cortex but are expected to perform worse on object and semantic representations. Neither type of method has been tested on subcortical representations, or on clinically relevant functions such as pain and emotion. Third, these methods do not include an explicit spatial model able to make inferences on the location or extent of activation, though a few recent studies have taken steps in this direction \citep{Nenning2017}. 

In this work, we focus on performing local registration of brain activation using subject-specific functional activation maps. These maps are the output of a series of voxel-wise models that quantify the activation in the BOLD signal that arises due to the stimulus of interest. 
They correspond to spatial maps of the regression coefficient corresponding to the predictor modeling the stimulus; for more detail see Section \ref{data description}. As these are coefficient estimates from a normal-error linear model that exhibit significant spatial correlation they are well-described using a Gaussian process. It is important to note that these models are fit after an initial anatomical registration has already been performed. Thus, all subject-specific activation maps should be in a common anatomical space, and all voxel-specific values directly comparable across subjects. However, as discussed above, there remains substantial individual variation in functional brain anatomy even after anatomical registration that causes this assumption to be invalid. To circumvent this issue, we seek to perform a secondary functional registration of the data into a common functional space. This can be accomplished by warping (i.e., spatially transforming) a set of floating maps (i.e., subject-specific activation maps) to some predefined reference map (i.e., a particular subject-specific activation map or a group-averaged map).  

In service of this goal, we propose a generalized Bayes approach that is capable of cropping the subject-specific activation maps to match certain pre-defined regions of interest (ROIs).  We adapt a flexible general loss function-based pseudo-likelihood approach to measure and correct for relative misalignment between the floating and reference maps.  The pseudo-likelihood is a representation of the similarity measure between the reference map and the warped floating map. The measure can be the sum of squared difference (SSD) or any other context-dependent choice. The parameters characterizing the misalignment are assigned prior distributions estimated by matching the features of the activation maps, which are commonly defined by local peaks. Furthermore, our work simultaneously models an intensity correction term which compensates for differences in the level of activation between the reference and floating maps. As the images are defined on a regular lattice, the discreteness leads to the need to utilize Gaussian process based Kriging interpolation techniques \citep{SBanerjee2014,cressie2015statistics} while realigning images. 
Finally, inference on the model parameters are based on the draws from the posterior distribution. We use the Markov Chain Monte Carlo (MCMC) algorithm to simulate from the posterior distributions.




This paper is organized as  follows. Section \ref{data description} describes the data, which is from an fMRI study of thermal pain.  In Section \ref{Methods} we introduce our generalized Bayes framework for performing functional image registration. In Section \ref{Simulation} we describe a simulation study used to evaluate the performance of the method, and in Section \ref{DataAnalysis} we apply the method to data from the thermal pain study. The paper concludes with a discussion.

\section{An fMRI Study of Thermal Pain } \label{data description}

The data are from an fMRI study of thermal pain; see \cite{woo2015distinct} for an in-depth discussion of the data set. In brief, $33$ healthy, right-handed subjects completed the study (age $27.9 \pm 9.0$ years, $22$ females). All subjects gave informed consent, and the Columbia University Institutional Review Board approved the study. 

Seven runs were administered to each subject during a single session, with each run consisting of between 58-75 trials. In our analysis the runs were concatenated and analyzed together. In each trial, thermal stimulations were delivered to the left inner forearm. Each stimulus lasted $12.5$s, with $3$s ramp-up and $2$s ramp-down periods and $7.5$s at the target temperature. In total stimuli were administered at six different temperatures, ranging from $44.3 - 49.3^\circ$C in increments of $1^\circ$C.   Each stimulus was followed by a $4.5 - 8.5$s pre-rating period, after which subjects rated their intensity of pain on a scale between $0 - 100$. 
Each trial ended with a $5 - 9$s resting period.

For each subject, 1845 images were acquired using a 3T Philips Achieva TX scanner at Columbia University. Structural images were acquired using high-resolution T1 spoiled gradient recall (SPGR) images.
Functional echo planar images (EPIs) were acquired with repetition time (TR) = $2000$ms, echo time (TE) = $20$ms, field of view = $224$mm, $64 \times 64$ matrix, $3 \times 3 \times 3$mm$^3$ voxels, $42$ interleaved slices, parallel imaging, and sensitivity encoding (SENSE) factor $1.5$. For each subject, structural images were co-registered to the mean functional image using the iterative mutual information-based algorithm in SPM8\footnote{Statistical Parametric Mapping, version 8; http://www.fil.ion.ucl.ac.uk/spm/}. Thereafter, images were normalized to Montreal Neurological Institute (MNI) space using SPM8's generative segment-and-normalize algorithm. Prior to preprocessing of functional images, the first four volumes were removed to allow for image intensity stabilization. Outliers were identified using the Mahalanobis distance for the matrix of slice-wise mean and standard deviation values. The functional images were corrected for differences in slice-timing, and motion corrected using SPM8. These images were warped to SPM's normative atlas using warping parameters estimated from co-registered high-resolution structural images, and smoothed with an $8$mm full width at half maximum (FWHM) Gaussian kernel. A high-pass filter of $180$s was applied to the time series data. 

For each subject, a voxel-wise general linear model (GLM) analysis was performed using SPM8. For each temperature, boxcar regressors convolved with the canonical hemodynamic response function \citep{lindquist2009modeling}, were constructed to model periods corresponding to the $12.5 \, s$ thermal stimulation and the $11 \,s$ rating periods. Other regressors of non-interest (i.e., nuisance variables) included (a) a run specific intercept; (b) linear drift across time within each run; (c) the six estimated head movement parameters (x, y, z, roll, pitch, and yaw), their mean-centered squares, their derivatives, and squared derivative for each run; (d) indicator vectors for outlier time points identified based on their multivariate distance from the other images in the sample; (e) indicator vectors for the first two images in each run; (f) signals from white matter and ventricle. 
For each temperature the estimated regression coefficients corresponding to that temperature specific regressor at each voxel were combined into a single activation map. Hence, for each subject we had six different brain maps depicting the functional response across the brain to each of the temperatures. 

In this work, we are primarily focused on the subject-specific mean across the six difference brain maps to analyze the inter-subject variability.  We use this data to perform functional alignment. In particular, we focus our analysis on a somatosensory region of the brain; Fig \ref{fig:realData:raw} shows {\color{black} a circular region of radius 15 voxels} across subjects. Within this region one can identify several local peaks within individual participants that conform to known somatosensory areas. However, there clearly exists substantial inter-subject variability in their exact location. 
The somatosensory region of the brain was chosen specifically because it is known to be related to pain \citep{bushnell1999pain, gustin2012pain, vierck2013role}. It is {\color{black} reliably} observed in pain studies as evidenced by meta-analyses of neuroimging pain studies (see, for example, NeuroSynth \citep{yarkoni2011large}). 

\section{Methods} \label{Methods}

The goal of image registration is to find a mapping from each voxel in one image (the floating map) to a corresponding position in another image (the reference map).
To formalize the problem, let $\Y  = (Y (s_1), Y (s_2), \dots, Y (s_V))^T$ and $\R = (R(s_1), R(s_2), \dots, R(s_V))^T$ be the floating map and reference map, respectively, located at the same set of voxel locations $\Sbf = \{s_1, s_2, \cdots, s_V\}$. This mapping involves operations (e.g., translation, scaling, rotations, or more complex non-linear transformations) described by a transformation operator $T$ on the floating map $\bY$ to align it with the reference $\bR$.  

Let $\w$ denote the set of parameters characterizing the transformation function $T := T(\cdot,\bw)$. Our goal is to estimate $T$ via $\bw$ with proper uncertainty quantification. Hence, we seek a Bayesian formulation of the registration problem that will allow obtaining credible intervals for the transformation parameters via posterior MCMC samples. A conventional approach for Bayesian treatment of such a registration problem would require likelihood from a generative model that represents the relationship between the floating and reference maps. One can consider a model of the form 
$\R \sim_d F(\bY(T),\theta)$, where $\bY(T)=(Y(T(s_1)),\ldots,Y(T(s_V))^T$.
This requires the specification of the full probability distribution $F$, parametrized via $\theta$. A distribution-free approach is less assumption-driven and more desirable. 
Also, such a probability model will typically posit the reference map $\R$ to be noisier than the floating map. For example, one can specify a generative model $F$ as \begin{equation}\label{eq:model}
\R(s) = b \Y(T (s, \mathbf{w})) + \e(s), \;\; \text{for} \; s \in \Sbf,
\end{equation}
where $\Y (T (s, \w))$ is the activation of the floating map at the transformed location $T(s, \w )$, the factor $b$ adjusts for the discrepancy of intensities between the floating and reference maps, and $\e(s)$ is the additive noise.  
It is evident from (\ref{eq:model}) that $\bR$ is modeled as noisier than $\Y$ due to the errors $\mathbf e$. This is  undesirable as in practice the observed floating map is expected to be equally or more noisy than the reference, and hence generative models of this form are likely to be misspecified.

Furthermore, the generative model approach becomes problematic when working with multiple floating maps $\bY_i$, as it creates multiple generative models for the common reference map $\bR$. 
Another possible model-based solution to the multiple registration problem would be to switch the roles of $\bY$ and $\bR$ in (\ref{eq:model}), expressing $\bY(s) = b\bR(T_{rv}(s,\bw')) + $error, where $T_{rv} := T_{rv}(\cdot,\bw')$ denotes the reverse transformation, mapping $\bR$ to $\bY$, parametrized by $\bw'$. This formulation easily extends to multiple images $\bY_i$,  naturally modeling all the floating maps using the common reference map $\bR$, and mitigates the issue of the reference map being modeled as more noisy. However, the model fit $\hat b \bR (T (s, \hat \bw))$ removes all subject-specific features that are not present in the reference.  Also this switched approach does not offer any direct way to register the floating image to the reference. To remedy this, one often leverages the notion of {\em inverse consistency}, jointly estimating both the forward and the reverse maps $T$ and $T_{rv}$ with the constraint that their composition is the identity operator. Such an approach can be seamlessly implemented when using two loss-functions, one characterizing similarity between $\bR$ and $\bY(T)$ and the other between $\bY$ and $\bR(T_{rv})$. However, it is challenging to make this approach amenable to a fully Bayesian treatment, as it would involve two different generative models $\bY | \bR(T_{rv})$ and $\bR | \bY(T)$ for the same pair of data $(\bR,\bY)$. As these properties are undesirable, alternative approaches must be explored to conduct Bayesian registration. 

\subsection{Generalized Bayes framework for registration} For the above mentioned reasons, we abandon a fully model-based formulation and instead adopt 
a more flexible general loss function based approach to measure and correct for relative misalignment of the individual map $\Y $ with respect to the reference map $\R$. We first consider a loss-function $\ell(\bR,\bY(T),\w)$ corresponding to the forward transformation $T$ 
where $\bY(T)=(Y(T(s_1),\ldots,Y(T(s_V))^T$ and $\w$ denotes all the parameters involved in the transformation operator $T$ and the loss function $\ell$. 

If $\pi(\w)$ denotes a prior on the parameters, then a posterior distribution given the loss function $\ell$ will be given by
\begin{equation}\label{eq:gibbspost}
    \pi(\w \;|\; data) = \frac{\exp\left(-\ell(\bR,\bY(T),\w)\right)\pi(\w)}{\int \exp\left(-\ell(\bR,\bY(T),\w)\right)\pi(\w)d\w}
\end{equation}

Posteriors of the form (\ref{eq:gibbspost}) that do not rely on a probability model for the data but simply a loss function are called Gibbs posteriors or generalized posteriors. They feature in early works of \cite{vovk1990aggregating,shawe1997pac,mcallester1999some,walker2001bayesian} and their posterior summaries are referred to as Laplace Type Estimators (LTE) in \cite{chernozhukov2003mcmc} who developed extensive theoretical results on posterior concentration, asymptotic normality and valid confidence intervals for such posteriors and estimators. \cite{bissiri2016general} argued that updates of the form (\ref{eq:gibbspost}) are the only coherent update of a prior belief $\pi$ given the data and the choice of the loss function $\ell$. Using Gibbs posteriors are becoming an increasingly common form of Bayesian inference, termed as {generalized Bayes} \citep{grunwald2020fast,fiksel2021generalized,rigon2020generalized}. 

The loss function approach is compatible with multiple registration to a common template. When working with multiple floating maps, the total loss is simply the sum of the losses for the individual maps. For example, registering $K$ images $\bY_i, i=1,\ldots,K$ to a common $\bR$ entails simple extension of the loss function to $\sum_{i=1}^K \ell(\bR,\bY_i(T),\w_i)$ where the parameters $\w_i$ can be image-specific or shared across images, i.e., $\w_i=\w$ depending on the context. Thus the generalized Bayes approach for registration, by dissociating loss-functions from likelihoods of generative models, avoids the incompatibility (discussed in Section \ref{Methods}) of the model-based approach to register multiple floating maps. 



We define the loss function using the sum-of-squared differences (SSD) between the warped image $\Y(T)$ and $\R$,   
\begin{equation}\label{eqn: loss}
\ell(\R,\Y, b, \phi, \w)= \frac{1}{\phi^2} \|\R - b\Y(T(\cdot,\w))\|^2_{L_2}. 
\end{equation}
The generalized-Bayes approach also seamlessly extends to satisfy the property of inverse-consistency introduced by \cite{Inverse-consistency1} and \cite{Inverse-consistency2}. For this purpose, we tweak the loss function as follows: 
\begin{equation}\label{eqn: loss-inverseConsistency}
\begin{array}{cc}
    \ell_{ic}(\Y, \R, b,b', \phi, \w, \w')= & \ell(\R,\Y, b, \phi, \w) + \ell(\Y,\R, b', \phi, \w') +  \\
     & \lambda_{rv} \|T_{rv}(T(\cdot))-Id(\cdot)\|_F.
\end{array}
\end{equation}
Here $\w'$ characterizes the reverse transformation map, $T_{rv}(\cdot, \w')$, from $\R$ to $\Y$ and $Id(\cdot)$ is the identity map. The scaling parameter $\phi$ controls the {\em learning rate} from the loss function (relative to the prior) in generalized Bayes and is assumed to be the same for the forward and reverse losses, thereby assigning equal weight to both. The last quantity in (\ref{eqn: loss-inverseConsistency}) is the inverse-consistency penalty that constrains the composition of forward and reverse transformation to be close to the identity map. The hyper-parameter $\lambda_{rv}$ balances the intensity loss and inverse-consistency penalty.
While there are many potential image similarity measures \citep{Crum2004}, we use SSD as it has been demonstrated as being appropriate for single-modal brain registration \citep{Simpson2012}. However, the general framework outlined here is agnostic to the choice of loss functions, and other losses like mutual information loss \citep{Viola1997, Thevenaz1998}, correlation-based losses \citep{Gruen1987}, 
can be used depending on the application.

To implement this approach, decisions need to be made regarding how spatial interpolation is performed on the discrete images to obtain $\bY(T)$ used by the loss function, choices of transformation operator, reference map, prior and hyper-parmeters. We discuss these decisions in detail below. 

\subsection{Kriging Interpolation}\label{sec:kriging}
It is important to note that updating the parameters via optimization, or sampling using the loss function (\ref{eqn: loss}), would require evaluating the warped images $\bY(T)$ for a given value of the transformation parameters $\bw$. As the activation maps $\bY$ are only observed on a predefined set of $V$ voxels $\mathcal V$, 
the warped images, $Y(T)$ at the set of locations $T(\mathcal V)$, are acquired through spatial interpolation. While any interpolator can be used for this purpose, we use the Kriging interpolator \citep{matheron1963principles,SBanerjee2014,cressie2015statistics} 
\begin{equation}\label{eqn: Kriging distribution}
    \widehat{\Y (T )} = \frac{\one^T \C^{-1} \Y}{\one^T \C^{-1} \one}  \one  +  \K ^T\C^{-1} \left(\Y  - \frac{\one^T \C^{-1} \Y}{\one^T \C^{-1} \one}  \one \right)
\end{equation}
where $\one$ is a vector of ones of length $V$ and for a given choice of a covariance kernel $C$, $\C$ and $\K$ are $V\times V$ matrices with $(i,j)^th$ entries $C(s_i,s_j)$ and $C(s_i,T(s_j))$ respectively. We use the popular exponential covariance kernel, i.e., $C(s_i,s_j)=\sigma^2\exp\{-\rho\|s_i - s_j\|$ for any $s_i,s_j$, although any other kernel can be used. The spatial parameters $\sigma $ and $\rho $ can be estimated within the Bayesian framework jointly with the other parameters, although increasingly, a priori estimation of these parameters, as we do here, is being adopted as the pragmatic strategy \citep{finley2019efficient}.

\subsection{Transformation Operator}\label{sec:transformationoperator}
The general framework for Bayesian registration proposed via (\ref{eq:gibbspost}) and (\ref{eqn: loss-inverseConsistency}) can use any family of transformations. We here use the similarity transformation \citep{Cederberg2001}, which consists of translation, scaling, and rotation. For the case of 2D images this can be expressed as follows: 
\begin{equation}\label{eqn: transformation operator}
    T (\mathbf{s}) = \mathbf{A} \mathbf{s} + \mathbf{\theta} , \mbox{ where } 
\mathbf{A } = \begin{pmatrix}
\cos(\omega ) & -\sin(\omega )\\
\sin(\omega ) & \cos(\omega )
\end{pmatrix}
\times
\begin{pmatrix}
\sigma_{x} & 0\\
0 & \sigma_{y}
\end{pmatrix},
\end{equation}
where $\mathbf{s} = (s_x, s_y)^T$,
$\mathbf{\theta}  = (\theta_{x},\theta_{y})^T \in \mathbb{R}^2$, $\omega  \in (-\frac{\pi}{2}, \frac{\pi}{2})$, $\sigma_x>0$, $\sigma_y>0$. 
The similarity transformation in (\ref{eqn: transformation operator}) is a special case of the general 2D affine transformation. To see this, we decompose the transformation matrix $\mathbf{A}$ as
\begin{small} \begin{equation}\label{eqn: decomposition A}
\begin{aligned}
    &\mathbf{A}= 
    \begin{pmatrix}
    A_{11} & A_{12}\\
    A_{21} & A_{22}
    \end{pmatrix} \\
    &= 
    \begin{pmatrix}
    \cos(\omega) & -\sin(\omega)\\
    \sin(\omega) & \cos(\omega)
    \end{pmatrix}
    \begin{pmatrix}
    1 & \frac{A_{12}\cos(\omega) + A_{22}\sin(\omega)}{A_{22}\cos(\omega)-A_{12}\sin(\omega)}\\
    0 & 1
    \end{pmatrix}
    \begin{pmatrix}
    \text{sign}(A_{11})\sqrt{A_{11}^2 + A_{21}^2} & 0\\
    0 & A_{22}\cos(\omega)-A_{12}\sin(\omega)
    \end{pmatrix},
\end{aligned}
\end{equation}
\end{small}
where $\omega = \arctan(A_{21}/A_{11})$.
We note that the expression above is the result of the QR decomposition. The first term is the rotation matrix. It is followed by the shearing matrix. The last term is the scaling matrix. The equivalence between Eqns. (\ref{eqn: transformation operator}) and (\ref{eqn: decomposition A}) can now be derived by adding constraints to eliminate the shearing effect, i.e., setting $\frac{A_{12}\cos(\omega) + A_{22}\sin(\omega)}{A_{22}\cos(\omega)-A_{12}\sin(\omega)}=0$, or equivalently $A_{11}A_{12} + A_{21}A_{22}=0$. Under this constraint, the decomposition can be simplified as 
 \begin{equation}
    \mathbf{A} = 
    \begin{pmatrix}
    A_{11} & A_{12}\\
    A_{21} & A_{22}
    \end{pmatrix}
    = 
    \begin{pmatrix}
    \cos(\omega) & -\sin(\omega)\\
    \sin(\omega) & \cos(\omega)
    \end{pmatrix}
    \begin{pmatrix}
    \text{sign}(A_{11})\sqrt{A_{11}^2 + A_{21}^2} & 0\\
    0 & A_{22}/\cos(\omega)
    \end{pmatrix}.
\end{equation}
Therefore, there exists a bijection mapping between $(\sigma_x, \sigma_y, \omega)^T$ and $\vect{A}$ under the constraint that $A_{11}A_{12} + A_{21}A_{22}=0$. Furthermore, if we are interested in the positive scalings, $\sigma_x>0$, $\sigma_y>0$, this can be achieved by setting $A_{11}>0$ and $A_{22}>0$. We will exploit this connection to formulate the prior for the transformation parameters in the next Section.

\subsection{Regularization Prior}\label{sec:priors}
There are several obstacles in optimizing for transformation parameters in the general registration  problem. First, it is an inherently ill-posed problem. A classic example where two distinct registration routines might lead to the same registration result was discussed in \cite{Fischer2008}, where they showed a subtle example that either translation or rotation can yield the same optima. Second, it is difficult to avoid unwanted local optima. Both of these difficulties can be addressed by including a regularization, or penalty, term. In this Section we adapt this idea by introducing regularization priors for our generalized Bayes framework. 

The generalized (Gibbs) posterior distribution of the warping parameters takes the form:
\begin{align}\label{eqn:posterior}
    p(T,T_{rv},\phi, b, b'| \Y, \R)\propto\exp{\left\{-\ell_{ic}(\Y, \R, b,b', \phi, \w, \w')\right\}}\pi(T,T_{rv}, \phi, b, b'),
\end{align}
where $\pi(T,T_{rv}, \phi, b, b')$ is the regularization prior. We specify 
$\pi(T,T_{rv}, \phi, b, b') =$ \\ $\pi(T| \phi)\pi(T_{rv}|\phi)\pi(b'|\phi)\pi(b| \phi)\pi(\phi)$ for choices of $\pi(T| \phi)$, $\pi(T_{rv}| \phi)$, $\pi(b| \phi)$  and $\pi(b'| \phi)$ which leads to the following form of the negative log-posterior, 
\begin{align}\label{eq:logpost}
    &\frac{1}{2\phi^2}\|\R - b\Y(T)\|_{L_2} ^2 + \frac{1}{2\phi^2}\|\Y - b\R(T_{rv})\|_{L_2} ^2 + \lambda_{rv} \|T_{rv}(T(\cdot))-Id(\cdot)\|_F \\ \nonumber
    &+\frac{\lambda_T}{\phi^2} \Reg(T) + \frac{\lambda_b}{\phi^2} (\log(b)-\log(b_0))^2 + \frac{\lambda_{T_{rv}}}{\phi^2} \Reg(T_{rv}) + \frac{\lambda_{b'}}{\phi^2} (\log(b')-\log(b'_0))^2,
\end{align}
where the regularizer $\Reg$ can be chosen in different ways. The terms $\lambda_b$ and $\lambda_T$ (and analogously, $\lambda_{b'}$ and $\lambda_{T_{rv}}$) are the tuning parameters, which control the trade-off between the fitting of the registration likelihood and the regularization priors. The factor $(\log(b)-\log(b_0))^2$ in the regularization part regularizes the intensity-correction term, and the term $b_0$ is one of the hyper-parameters whose choice will be described in the next Section. 
To gain intuition on the necessity of this term, 
without penalizing values of $b$ away from some $b_0$, the MCMC may become stuck in loss function valleys centered around a local minima. 
For example, around $b=0$, the loss function will be flat for all values of the transformation parameters $w$. The prior for $b$ leading to (\ref{eq:logpost}) is given by $\log(b) | \phi \sim N(\log(b_0), ~ \frac{\phi^2}{\lambda_b})$. 
The regularization of $T$,  $\Reg(T)$, is taken to measure the distance between the coordinates transformed by the proposed $T$ and the some initial warping map $T_0$ defined by hyper-parameters $A_0$ and $\mathbf{\theta}_0$. 
The selection of these hyper-parameters is critical to ensure identifiability of the registration, and is discussed in detail in Section \ref{sec:features}. 
Formally, the regularizer of $T$ is represented in the following way: 
\begin{equation}
    \Reg(T) = \sum_{\s\in \mathcal{R}}\|T(\s) - T_0(\s)\|_{L_2}^2 = \sum_{\s\in \mathcal{R}}\|(A-A_0)\s + \mathbf{\theta} - \mathbf{\theta}_0\|^2_{L_2}.
\end{equation}
 
The expression of the regularizer can be further simplified as:
\begin{align*}
    \Reg(T) &=\text{trace}\{[\s_x ~ \s_y ~ \one] (M-M_0) (M-M_0)^T [\s_x ~ \s_y ~ \one]^T\}\\
    &= \text{trace}\{(M-M_0)^T \Sigma_s (M-M_0)\},
\end{align*}
where 
$M = [A ~ \mathbf{\theta}]^T = 
\begin{pmatrix}
A_{11} & A_{12} & \theta_x\\
A_{21} & A_{22} & \theta_y
\end{pmatrix}^T$, 
$M_0 = [A_0 ~ \mathbf{\theta}_0]^T$, and
$\Sigma_s = $
$\begin{pmatrix}
\s_x^T\s_x & \s_x^T\s_y & \s_x^T\one\\
\s_x^T\s_y & \s_y^T\s_y & \s_y^T\one\\
\s_x^T\one & \s_y^T\one & \one^T\one\\
\end{pmatrix}$.

Hence, using the prior $\vect{M} | \phi \sim N(\vect{M_0}, ~ \frac{\phi^2}{\lambda_T} \I\otimes\Sigma_s^{-1})$, where $\otimes$ denotes the Kronecker product and $\I \in \mathbb{R}^{2\times 2}$ the identity matrix, ensures that the Gibbs posterior is of the form (\ref{eq:logpost}). Note that, $M$ has $6$ parameters and our transformation is only defined by $5$ parameters $\omega,\sigma_x,\sigma_y,\theta_x,\theta_y$. As discussed in Section \ref{sec:transformationoperator}, there is a one-to-one relationship between our parametrizations and the general 2D affine transformation parametrized by $M$ under the constraint $A_{11}A_{12} + A_{21}A_{22}=0$. Hence, we only use the  multivariate normal prior for the 3-dimensional subset of $A$ along with $\mathbf{\theta}$ to correspond to the priors on the rotation, scaling and translation parameters. 

Similarly, we define the regularizer of $T_{rv}$ as
\begin{align}
    \Reg(T_{rv}) = \text{trace}\{(M'-M'_0)^T \Sigma_s (M'-M'_0)\},
\end{align}
where $M' = [A' ~ \mathbf{\theta}']^T$, $T_{rv}(s) = A'\s+\theta'$. $M'_0 = [A'_0 ~ \mathbf{\theta}'_0]^T$ is the prior transformation parameters of $T_{rv}$. To meet the requirement of inverse-consistency, the prior mapping, $T_{rv0}$ is constrained to 
the inverse of $T_0$. In other words, we have $A'_0(A_0\s+\theta_0)+\theta'_0 = \s$.

Finally, we use the standard reference prior for the scale parameter $\phi$, i.e., $\pi(\phi) \propto 1/\phi^2$. 
Combining all the pieces in the generalized Bayes framework, the Gibbs posterior distribution of the warping parameters has the following form, 
\begin{align}\label{eqn:posterior_inv}
    &p(\mathbf{\theta}, \mathbf{\sigma}, \omega)\propto\\
    &\phi^{-2V-14}\exp\Bigg\{ -\frac{1}{2\phi^2}\|\R - b\Y(T)\|_{L_2}^2 -\frac{1}{2\phi^2}\|\Y - b'\R(T_{ic})\|_{L_2}^2 \nonumber\\
    & -\frac{\lambda_b}{2\phi^2}((\log(b)-\log(b_0))^2)  -\frac{\lambda_T }{2\phi^2}\text{trace}\{(M-M_0)^T \Sigma_s (M-M_0)\}\nonumber \\
    & -\frac{\lambda_{b'}}{2\phi^2}((\log(b')-\log(b'_0))^2) -\frac{\lambda_{T_{rv}}}{2\phi^2}\text{trace}\{(M'-M'_0)^T \Sigma_s (M'-M'_0)\}\Bigg\}.\nonumber
\end{align}
Notice that we have dropped the explicit penalty for inverse-consistency, as it is embedded in the choice of prior transformations. 


\subsection{Corresponding Feature-based Prior Estimation}\label{sec:features}
Informed selection of the regularization hyper-parameters $b_0$ and $M_0$ is critical to ensure identifiability of the registration problem. In this Section we outline an algorithm for prior (hyper-parameter) estimation based on corresponding features. The loss function (\ref{eqn: loss}) is solely based on the intensity information of the images, and does not use the landmark information of activation maps which is an alternate branch in the image registration. The prior (hyper-parameters) of the transformation can be estimated by incorporating information about the landmarks. This approach does not encounter computational burden with a moderate number of landmarks. 
In this work, we define the landmarks as the regional peaks (maxima) in clusters consisting of eight-adjacent voxels. 

Let $\bfp^R = \{s_i^R \in \mathbb{R}^2| i = 1, \cdots, N_R\}$ and $\bfp^Y = \{s_i^Y \in \mathbb{R}^2| i = 1, \cdots, N_Y\}$ be the landmarks of reference map and the floating map, respectively. $N_R$ and $N_Y$ are the number of landmarks in the reference map and floating map, respectively. 
As a first step, the coarse sub-region of interest of the reference map is determined manually based on scientific beliefs. This sub-region contains a subset of landmarks, $\hat\bfp^R \in \bfp^R$, of size 
denoted by $\hat N_R$. We aim to find the corresponding subset of the landmarks $\hat \bfp^Y$ from $\bfp^Y$. 
Many conventional methods, including Geometric Hashing \citep{Mian2006} and Iterative Closest Point \citep{Yang1992, Besl1992}, work for point set matching. However, these methods fail to capture the intensity features of the maps.
In our application, as the size of the set of landmarks are relatively small (i.e., $N_Y<20$ and $\hat N_R < 5$), it is computationally efficient to apply the Brute Force Search method that tries all possible $\hat N_R$ dimensional subsets of $\bfp^Y$ to determine the matching landmarks.

The similarity transformation can be estimated using Procrustes analysis. Here we apply a modified version of Procrustes analysis, namely ABC Procrustean Algorithm \citep{Awange2008}. This version is preferable to the ordinary Procrustes algorithm as it is capable of finding the distinct scale parameters in each axial direction as opposed to the estimation of uniform scales. 

The objective function used to search for correspondence is selected as the sum of squared distance between the query landmarks and the transformed candidates. 
Again, we need to add a constraint term to address the ill-posed registration problem, as discussed in the beginning of this Section. Formally, the mechanism can be formulated as follows,
\begin{align}\label{eqn: main corresponding mechanism}
     &||\hat\bfp^R - T_\alpha(\hat\bfp^Y)||^2 + ||\hat\bfp^Y - T^{-1}_\alpha(\hat\bfp^R)||^2 \leq 2d \\
     \text{subject to}\;\;\;\; &S(\hat\bfp^Y - T_\alpha(\hat\bfp^Y)) + S(\hat\bfp^R - T^{-1}_\alpha(\hat\bfp^R))  < \alpha, \nonumber
\end{align}
using the transformation operator $T_\alpha(\s) =  \A_\alpha \s + \mathbf{\theta}_\alpha$. 

The fact that the activation maps are discrete realizations of continuous maps motivates the use of the distance less than the threshold $d$, instead of minimizing the distance between landmarks. The discreteness of a map results in the bias of the coordinates of landmarks, as the underlying location of landmarks can be anywhere within a pixel. We suggest to take $d = \hat N_R$ which relaxes the landmark within a voxel. The objective function creates a set of candidates of correspondence, and we choose the optimal correspondence based on the intensity loss. 

The constraint $S$ is motivated by diffusion registration \citep{Fischer2003} and has been applied to affine models by \cite{Chumchob2009}. By defining a function $D\colon\mathbb{R}^2 \rightarrow \mathbb{R}^2$ with $D(\bfp) = (D_1(\bfp), D_2(\bfp))^T= \bfp - T(\bfp)$, it can be expressed as follows:
\begin{align}\label{eqn: constraints}
     S(\bfp - T(\bfp)) = \frac{1}{2} \int ( |\nabla D_1(\bfp)|^2 + |\nabla D_2(\bfp)|^2) d\bfp = \|\I-\A||_F^2,
\end{align}
where $\|\cdot\|_F$ denotes the Frobenius norm, $\I$ is the identity matrix, $\A$ is defined in Eq. (\ref{eqn: transformation operator}). 
The other application of the constraint defined in Eq. (\ref{eqn: constraints}) is to determine the orientation of $\hat \bfp^Y$. For example, in the case of $\hat N^R = 3$, $\hat \bfp^Y = \{s_1^Y, s_2^Y, s_3^Y\}$, there are six possible correspondences to $\hat \bfp^R$, and the transformation can be derived for each of them. The mismatched correspondence may result in excessive transformation, which can cause redundant rotation and scaling. Therefore, the one minimizing $S(\hat\bfp^Y - T_\alpha(\hat\bfp^Y))$ is used to determine the orientation of $\hat \bfp^Y$.

The threshold $\alpha$ in Eq. \ref{eqn: main corresponding mechanism} is chosen automatically using the Photometric Error Criterion (PEC) \citep{Brunet2010}. For each $\alpha$, the optimization problem in Eq. (\ref{eqn: main corresponding mechanism}) is used to determine the corresponding features as well as the associated transformation. To evaluate the hyper-parameter $\alpha$, the criterion takes advantage of intensity information, which is defined as:
\begin{equation}\label{eqn: PEC_inv}
    \mathcal{C}(\alpha) = \frac{1}{|\mathcal{G}|}\sum_{\bfp\in\mathcal{G}}|| \R(\bfp)-b\Y(T_\alpha(\bfp))||^2 +\frac{1}{|\mathcal{G'}|}\sum_{\bfp\in\mathcal{G'}}|| \Y(\bfp)-b'\R(T^{-1}_\alpha(\bfp))||^2.
\end{equation}
Here $\mathcal{G}$ and $\mathcal{G'}$ are the regions of interest in $\R$ and $\Y$ respectively, and $|\cdot|$ defines the size of the region. The discrepancy of scale between the target map and the reference map is corrected by a scaling factor $b$. Note that $b$ can be estimated as the regression coefficient of $\R(\bfp)$ against $\Y(T_\alpha(\bfp))$ without an intercept as the criterion in Eq. ($\ref{eqn: PEC_inv}$) is denoted as the mean squared error of the regression.

Eq. $\ref{eqn: PEC_inv}$ is the summation of mean squared error of regressing $\R$ against $\Y(T_\alpha)$ and $\Y$ against $\R(T^{-1}_\alpha)$, where $b'$ is the regression coefficient. 
This estimate $b_0$ and $b'_0$ is used as a hyper-parameter in the prior of $b$ and $b'$ in Section \ref{sec:prior}.

We may further define the robust photometric error criterion (RPEC) by making a heavy-tailed distribution assumption on the residuals, for instance, the Student T distribution with a small degree freedom. The selected $\alpha$ minimizes the criterion in Eq. (\ref{eqn: PEC_inv}), and produces the feature correspondence and the estimation of transformation $M_0$ and $M'_0$ which will be considered as the hyper-parameter in the prior for $M$ and $M'$ respectively in Section \label{sec:prior}. We summarize the full steps of prior estimation in Algorithm (\ref{algorithm: priorSelection}).

\subsection{Choice of Reference Map}

A critical decision when performing registration is deciding on an appropriate reference map. One option is to use the group mean map across all floating maps. However, this will generally make the reference map blurry and the estimated transformation largely variant. In our application, the reference subject is chosen based on scientific knowledge. For example, researchers may choose the functional map that best represents the areas of interest to be the reference. 

It is worth noting that instead of aligning the whole brain functional map, we instead focus on estimating the transformation with respect to local regions of interest. In this way, estimation will not be unduly influenced by other nuisance regions that do not contain signal of interest. The query map cropped by a rectangle bounding box from the reference map describes the local area of interest; see Figure (\ref{fig:simulation_raw}) for an example. The transformed bounding boxes crop the floating maps to match and register to the query map. As the affine transformation we focus on is linear, the variability of bounding boxes on floating maps is equivalent to the variability of registration.

\subsection{Model Fitting}

The inference on the parameters is based on the draws from the posterior distribution. We use the Markov Chain Monte Carlo (MCMC) algorithm offered in the Stan software with default settings to simulate from the posterior distributions. Stan \citep{Stan2018} is implemented with the No-U-Turn sampler (NUTS) \citep{Hoffman2014}. We refer readers interested in details of the algorithm to the original work. Three distinct chains are fitted and initialized at values close to the prior estimates in order to accelerate the model fitting. 

The only hyper-parameters that are not determined in Section \ref{sec:features} are $\lambda_b$, $\lambda_T$, $\lambda_{b'}$ and $\lambda_{T_{rv}}$, which control the amount of regularization. It is common to choose such hyper-parameters using cross-validation. However the use of cross-validation embedded into MCMC runs is computationally intensive, and it requires a large amount of parallel computation to cover all possible splits of the data. \cite{Vehtari2017} proposed an approach for approximating leave-one-out cross-validation using Pareto smoothed importance sampling (PSIS-LOO) which can be easily computed. Compared to the Watanabe-Akaike information (WAIC; \cite{Watanabe2010}), we prefer PSIS-LOO as it is more robust in the finite case with weak priors or influential observations. 

Along with the PSIS, we also calculate the potential scale reduction statistic \citep{Gelman1992}, $\hat R$, as a metric of sampling convergence. Starting with a relatively high value of the hyper-parameters, we gradually fit the model with decreasing $\lambda$ until $\hat R < 1.01$. This step keeps the model away from  unwanted local modes. For each choice of ($\lambda_b$, $\lambda_T$  $\lambda_{b'}$, $\lambda_{T_{rv}}$) we compute PSIS and select the one minimizing PSIS.

The main source of computational complexity is from the Kriging interpolation, described in Section (\ref{sec:kriging}). The computational complexity is in the order of $O(V^3)$. While it is computationally manageable in our experiment due to the relative small $V$, the computation may be intractable for large $V$. In the case of large $V$, we recommend to apply the nearest-neighbor approximation introduced by \cite{Datta2016}, which will efficiently approximate the Kriging interpolation in the order of $O(Vm^3)$ for neighborhood size of $m<<V$.  As we are independently registering each subject's map, it is completely parallelizable across subjects. 

The models were fit in parallel across the choice of regularization parameters. Computations were carried out on a high-performance computing (HPC) cluster with an AMD EPYC 7702 64-core processor at 2.0GHz. We ran 10000 sampling iterations and set the initial 2000 iterations as the burn-in stage. The average computing time was 10 minutes for the burn-in stage and 20 minutes for the sampling stage. STAN has in-built diagnostics (e.g., the Gelman-Rubin potential scale reduction factor) that were used for determining convergence.
{\color{black} A software implementation of our method can be found at \url{https://github.com/gqwang1001/BayesianFunctionalRegistrationOfFMRIMaps}.}

\section{Simulation} \label{Simulation}

To illustrate the validity of our method we perform a simulation study using maps synthesized from the real data. The left panel of Figure \ref{fig:simulation_raw}A shows an example functional activation map. The query part of the map is manually cropped and shown within the dashed rectangle and in panel B. The floating map, shown in panel C, is generated by warping the reference map with the similarity transformation (including translation, scaling and rotation). The underlying parameters are translations $\theta = (2, -5)$, scalings $\sigma = (0.8, 1.2)$, and rotation $\omega = \pi/12$. White noise, $10^{-5}\epsilon$ with $\epsilon \sim N(0,1)$, is subsequently added to the map. 

We compare the learned prior information with the posterior in Figure \ref{fig:simulation_results}A, where the region surrounded by the dashed box represents the prior information and the solid box crops the matched region transformed by the posterior parameters. The boxes are generated by transforming the border box of the reference map using the prior estimates and median of posterior samples, respectively, of the transformation parameters. 

The models were fit using a sequence of regularization parameters ($\lambda_b$, $\lambda_T$, $\lambda_{b'}$, $\lambda_{T_{rv}}$) in Eq. (\ref{eqn:posterior}). The expected log predictive density (ELPD) was estimated on each model. Figure \ref{fig:simulation_results}B plots the -2 $\times$ ELPD which makes it on the deviance scale. The samples produced by the model with the lowest -2$\times$ ELPD were used to estimate the posterior distribution. The uncertainty of transformation can be demonstrated as posterior distribution of the translation, scaling and rotation parameters which are shown on Figure \ref{fig:simulation_densityPlot} with the truth superimposed as the dashed vertical line, where we can see the posterior samples successfully recover the truth. We further ran 100 simulations with different underlying transformation. Figure \ref{fig:simulation_100Comparison} shows the plot of posterior means against the true transformation. We could see the method recovered the truth successfully in each case.

To explore the models performance under a mis-specified model, the map is further deformed nonlinearly by elastic distortions, following the steps described in \cite{Simard2003GenerateSimulation}. First, the deformation field, $\Delta x$ and $\Delta y$ on $x$ and $y$ directions respectively, are generated from the uniform distribution, $\text{Unif}(-1, 1)$, at each voxel location. We then convolve the deformation field with the Gaussian filter with standard deviation $\sigma_d = 4$.  It is subsequently normalized and multiplied by a scaling factor $\alpha_d = 60$. 
Figure \ref{fig:nonlinear_simulation_100Comparison} shows the posterior mean plotted against the true affine parameters. Similar to the results shown in Figure \ref{fig:simulation_100Comparison}, the method successfully recovers the truth, while the deviance is larger due to the presence of nonlinear generalization.


\section{Real Data Results} \label{DataAnalysis}

We use the data described in Section \ref{data description} to perform functional alignment. Figure \ref{fig:realData:raw} shows functional activation maps of the region of interest across the $33$ subjects. Note that within this region there are multiple peaks. However, there also exists substantial inter-subject variability in their exact location. For example, comparing subjects 14 and 20 suggests a potential transformation consisting of rotation and shifting. Each of these functional activation maps are processed through the adaptive thresholding introduced by \cite{Bradley2007}. The landmarks are then identified as the regional maxima of the neighborhood of 8 adjacent locations. 

The same reference map is utilized as described in the simulation; see Figure \ref{fig:simulation_raw}. This particular map was chosen as it was determined to best represent the area of interest. 
The subject specific maps are shown in Figure \ref{fig:realData:prior}, and the estimated corresponding areas are indicated using the dashed boxes. Figure \ref{fig:realData:posterior} shows the matching area warped by the posterior mean of transformation, along with the 95\% credible region of the transformations are superimposed in the gray-shaded region. The credible region of transformation is found using the density-based clustering algorithm (DBSCAN), introduced by \cite{Ester1996}. The DBSCAN algorithm is the unsupervised clustering algorithm that separates the clusters based on the density. We refer readers to the original works for further details. To apply DBSCAN, we specify the epsilon region with at least 5 samples and a grid search of the size of the epsilon neighborhood to assure the only cluster captures about 95\% of the posterior samples.
The shaded region in the figure is then obtained by warping the searching box with each posterior samples of the transformation. 
The inter-subject variability in registration can be visualized using the estimated credible regions. For instance, subject 21 exhibits a high degree of uncertainty, while subject 14 exhibits relatively low uncertainty.
For subject 6, the estimated area falls outside the activation map. This is caused by the fact that we are analyzing a subset of the data corresponding to a circular `searchlight'. For this subject the best corresponding features lies on the edge of the searchlight, leading to a situation where the corresponding subregion falls outside of the searchlight range.

We contrast the results to those obtained using a standard non-linear deformation-based method. We applied the pipeline implemented in the image processing toolbox in MATLAB (version R2020b). 
Affine registration was first performed between each subject's activation map and the reference map, followed by diffeomorphic demons registration \citep{VERCAUTEREN2009Demons}. Figure \ref{fig:demonsResults} demonstrates the warping results. Seven out of thirty-three subjects, including subject 3, 5, 6, 12, 19, 25, 27,  failed to  register to the reference map. This is due to the diffuse activation maps. In addition, the standard approach was unable to provide the types of uncertainty measures of the registration that our approach provides. 

To evaluate the performance of the functional alignment procedure, we performed a second-level (across-subjects) analysis to test for significant temperature related effects. Figure \ref{fig:realData:group analysis} shows results comparing our approach to the standard approach, which first spatially registers images to a common template and then identifies activated regions, and the diffeomorphic demons approach.
Note that all approaches perform an initial anatomical registration into a common anatomical space and then compute activation maps. The difference lies in that our proposed approach and the demons approach thereafter perform a secondary functional registration of the derived activation maps. 
Studying the voxel-wise group means and standard deviations, as well as t-statistics show increased similarity in the activation profiles across subjects. After functional alignment, three peaks corresponding to posterior insula, SII, and opercular are readily apparent. In addition, the variation is decreased and more focused on the three regions-of-interest. This indicates that while the locations of the peaks are now consistent across subjects, the inter-subject differences in intensity remain. This is important for testing for brain-behavior correlations across subjects. Finally, the t-statistic is higher, indicating increased sensitivity for group-level inference. 
The results obtained using the demons algorithm show a similar pattern as seen using our approach. However, the t-statistic is significantly lower, indicating a lower sensitivity for group-level inference. 
Together, the results show the benefits of our functional registration approach.

\section{Discussion}

In this work, we propose a framework for reducing misalignment across individuals in functional brain systems. Our approach has the ability to register local brain features, defined as local peaks in the activation maps, providing a means to align functional features prior to inter-subject statistical analysis. Our generalized Bayes framework combines intensity-based information through the loss function and the feature-based information through the prior estimation, allowing us to conduct inference on the transformation via the posterior samples. In an application to fMRI data from a study of thermal pain, we found that the registration contributes to increased sensitivity for group-level inference. 

In its current implementation our approach has some limitations. First, in the case that significant non-linear differences between activation maps change the relative position between landmarks, our proposed method will likely fail. Using a more general class of transformations could address this issue.
Second, the choice of cropping boxes can affect the performance of registration. For example, in cases where the cropping box covers redundant area (e.g., including extra peak activation voxels that does not match the floating maps), the
performance of the proposed method may suffer due to the outlier region.
Third, we currently choose the reference image subjectively or based on the results of a meta-analysis. Therefore, in order to apply the method researchers need to check to ensure that the chosen reference pattern is shared across subjects. The proposed method can encounter difficult situations when the reference map is poorly defined. In future work, we will extend our approach to model the target image as a latent map to be estimated in a data-driven manner.

In this paper we have presented a 2-dimensional implementation of our proposed approach. However, we note that a 3-dimensional version could be constructed in an analogous manner. We began with the 2D version for illustration purposes, as it allows us to present a simplified description of the approach. In practice when applying the proposed approach to 3D data we would recommend a 3D implementation.

Our approach uses functional registration of activation maps to  account for residual differences in brain anatomy after spatial registration to MNI space has been performed. Once these functional transformations have been derived, it is possible to apply these transformations to the original 4D fMRI data and use these to compute functional connectivity. In addition, if the transformations are capturing true local differences in functional topology these transformations should carry over to other datasets collected on the same subject. For example, one could use activation maps from a task-fMRI to derive local transformations and then  apply these same transformation to resting-state data, which in turn can be used to compute functional connectivity. Another potential application of our proposed approach is predictive modeling. Here activation maps are  predictors used to model some behavioral or clinical outcome. Functional registration promises to align features in the model which can help improve prediction accuracy.
In future work we will seek to explore how functional registration can help improve population-level models of the functional brain representations that can be used to predict behavior, performance, clinical status and prognosis. We hypothesize that due to functional misalignment, features used in predictive models may not be properly aligned, negatively impacting predictive accuracy. Using our approach as a preliminary feature alignment step could potentially rectify this shortcoming.

Furthermore, as our data experiment measures multiple brain maps at six different temperatures, in future work we will model the within-subject variability using the registration and explore its association with other effects such as temperature. This will allow us to explore how the spatial extent of brain activation changes as a function of stimulus intensity. This provides a means to move past simply looking at spatial intensity in a specific voxel as a measure of brain activation. Another possible extension is evaluating the multilevel variability through the inference on the within-subject and between-subject transformation.
\bigskip

\section*{Acknowledgments}
The work presented in this paper was supported in part by NIH grants R01 EB016061 and R01 EB026549 from the National Institute of Biomedical Imaging and Bioengineering and R01 MH116026 from the National Institute of Mental Health.

\newpage

\bibliographystyle{apalike}
\bibliography{BayesianRegistration.bib}  

\newpage
\begin{algorithm}[H]
\label{algorithm: priorSelection}
Input: reference image $\R$, reference landmarks $\bfp^R$, floating image $\bY$, floating landmarks $\bfp^Y$, bound for distance between landmarks $d$, bound for tuning parameter $\alpha^{max}$.\\
Construct the set $\mathcal{S}^p = \{ \hat \bfp^Y \in \bfp^Y : | \hat \bfp^Y| = |\bfp^R|\}$,\\
Construct the corresponding transformation set $\mathcal{S}^T = \{\hat T : \hat T = \text{argmin} (\|\bfp^R - T(\hat \bfp^Y)\|) , \text{for}\;\; \hat\bfp^Y\in \mathcal{S}^p\}$.\\
\For{each $\alpha \in (0, \alpha^{max})$}{
    Find the subset $\hat{\mathcal{S}}^p_\alpha \subseteq \mathcal{S}^p$, such that $S(\hat\bfp^Y - T_\alpha(\hat\bfp^Y)) + S(\hat\bfp^R - T^{-1}_\alpha(\hat\bfp^R))  < \alpha \;\; \text{and}\;\; ||\hat\bfp^R - T_\alpha(\hat\bfp^Y)||^2 + ||\hat\bfp^Y - T^{-1}_\alpha(\hat\bfp^R)||^2 \leq 2d\}$\\
    Collect the corresponding transformation subset $\hat{\mathcal{S}}^T_\alpha \subseteq \mathcal{S}^T$,\\
    Compute the metric of $\alpha$, $\mathcal{C}(\alpha)$ defined in eqn. (\ref{eqn: PEC_inv}).
}
Output: choose $\alpha$ that minimizes $\mathcal{C}(\alpha)$. The corresponding $T_\alpha$ is the proposed prior. 
\caption{Corresponding Feature-based Prior (hyper-parameter) Estimation}
\end{algorithm}

\include{Figures}
\include{supplement}

\end{document}

%% file: Figures.tex
\begin{figure}
    \centering
    \includegraphics[scale = .5]{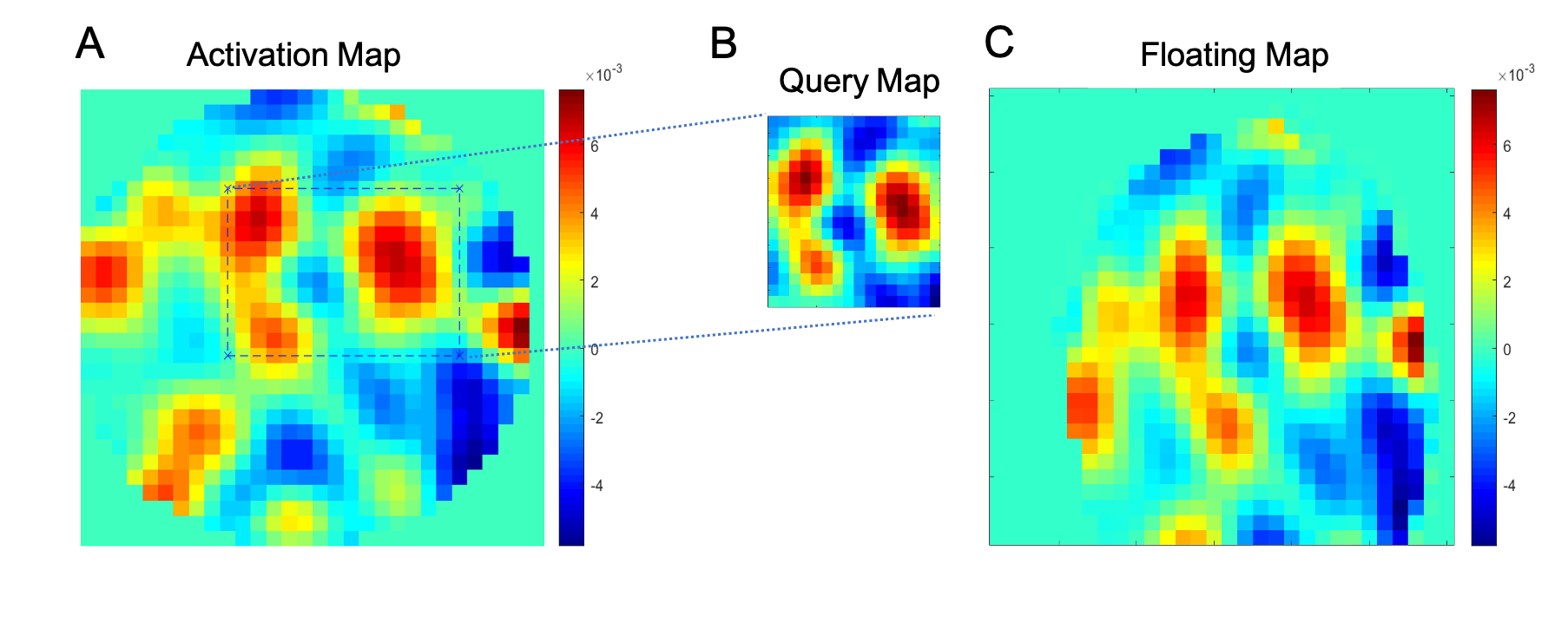}
    \caption{
    (A) An example of a functional activation map. When used as a reference map, a query map indicating landmarks of interest, is manually determined and shown within the dashed rectangle.
    (B) The query map extracted from the reference map. (C) The target map generated by warping the reference map.
    }
    \label{fig:simulation_raw}
\end{figure}



\begin{figure}
    \centering
    \includegraphics[scale = .7]{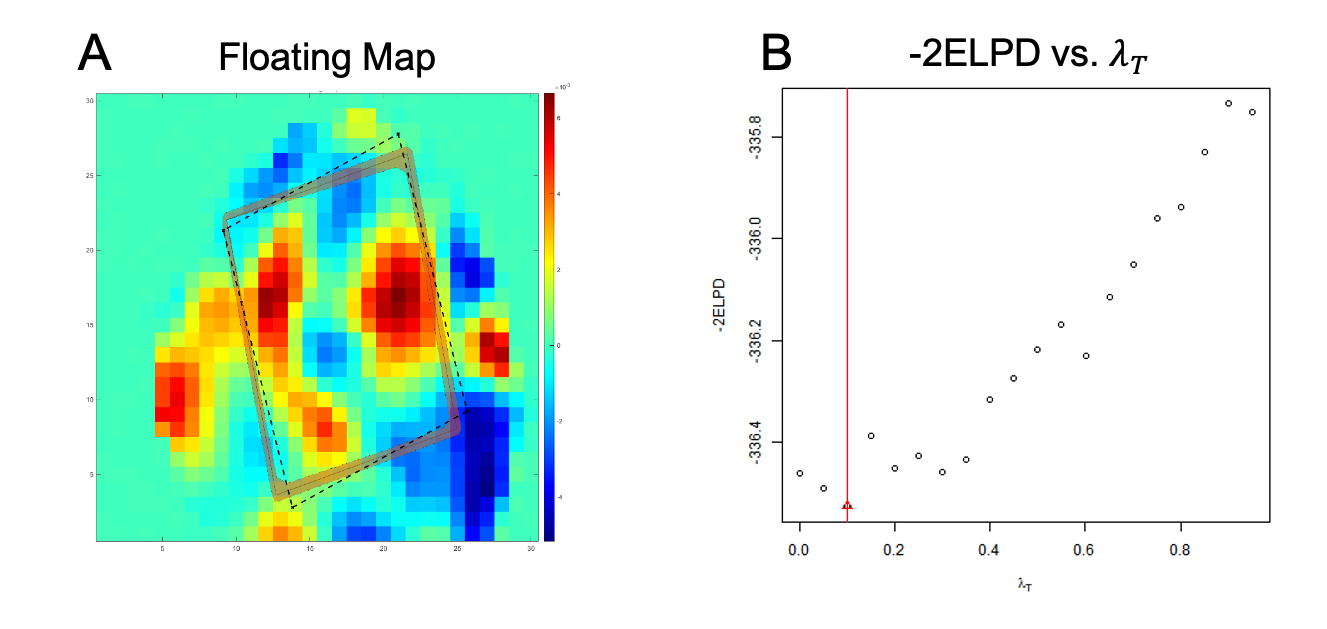}
    \caption{
    (A) The cropping box comparing the prior information (dashed box) with the posterior (solid box). (B) A plot of -2 $\times$ expected log predictive density (ELPD) versus the regularization parameter $\lambda_T$ given a fixed $\lambda_b$. The vertical bar (red) shows the selected $\lambda_T$, which reaches the minimum of -2$\times$ELPD.}
    \label{fig:simulation_results}
\end{figure}



\begin{figure}
    \centering
    \includegraphics[scale = .8]{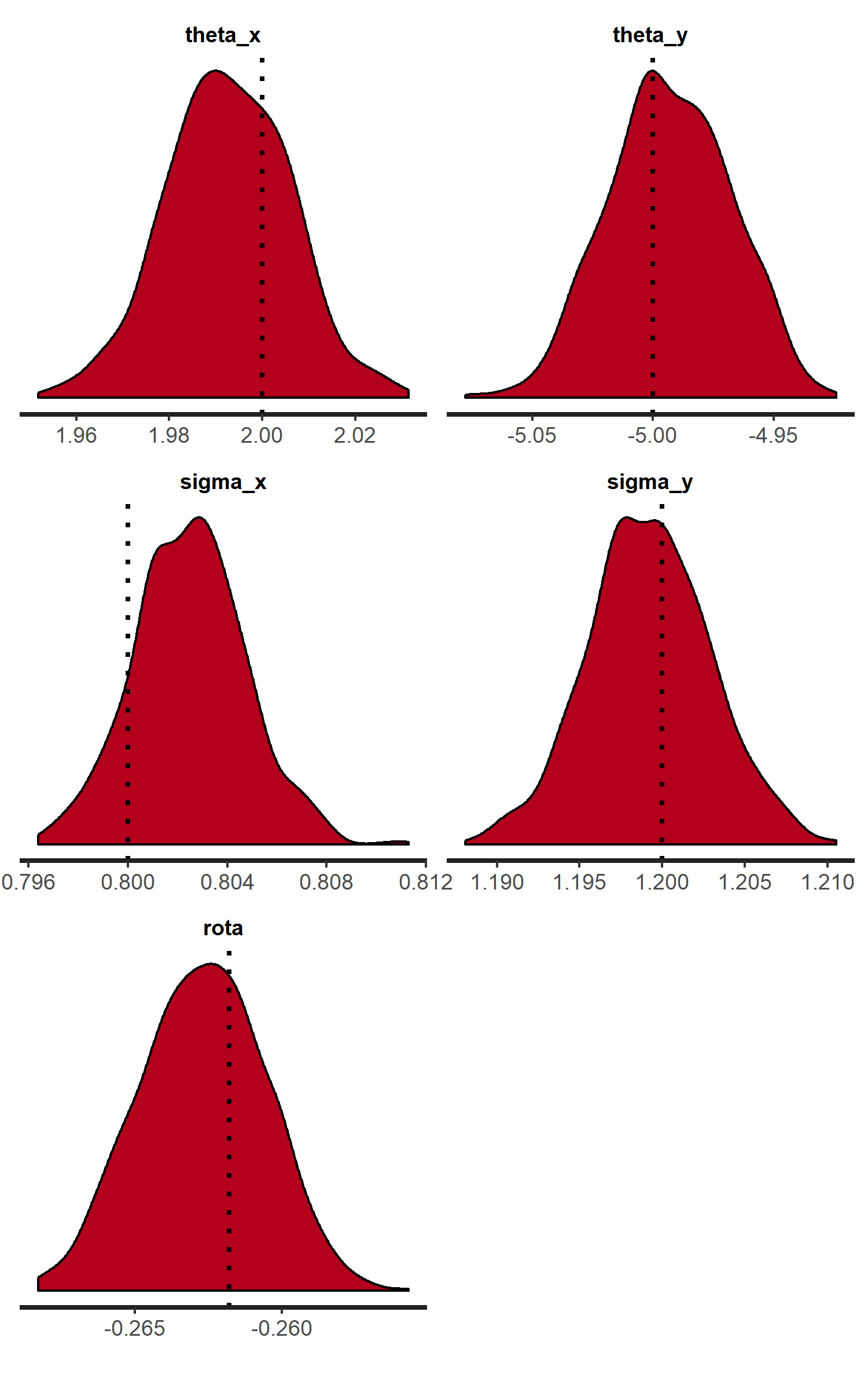}
    \caption{The estimated posterior density plot for each of the transformation parameters: translations (first row), scalings (second row), and rotations (third row). Vertical bars indicate the true parameter values.}
    \label{fig:simulation_densityPlot}
\end{figure}

\begin{figure}
    \centering
    \includegraphics[scale = .8]{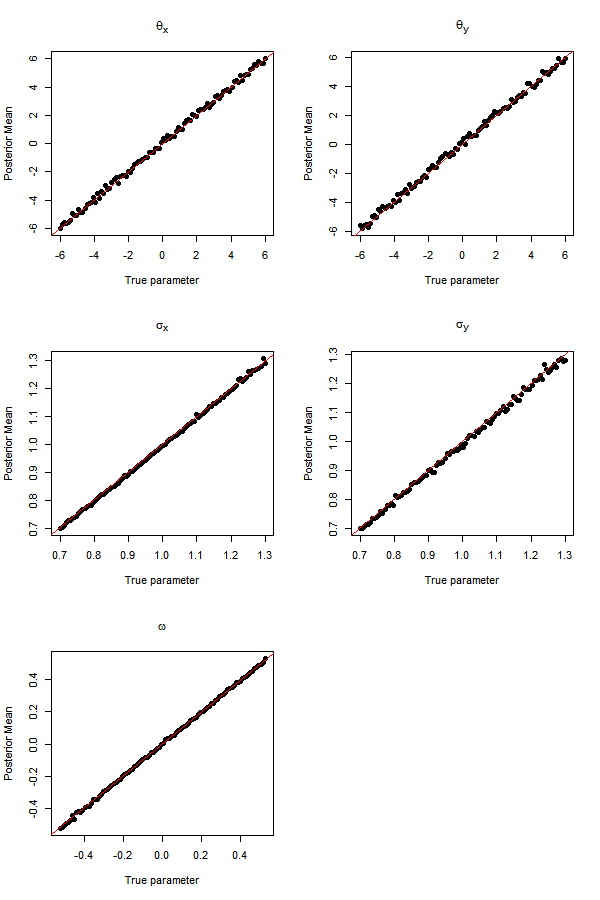}
    \caption{The posterior means plotted against the true parameters for each transformation parameter: translations (first row), scalings (second row), and rotations (third row).}
    \label{fig:simulation_100Comparison}
\end{figure}


\begin{figure}
    \centering
    \includegraphics[scale = .55]{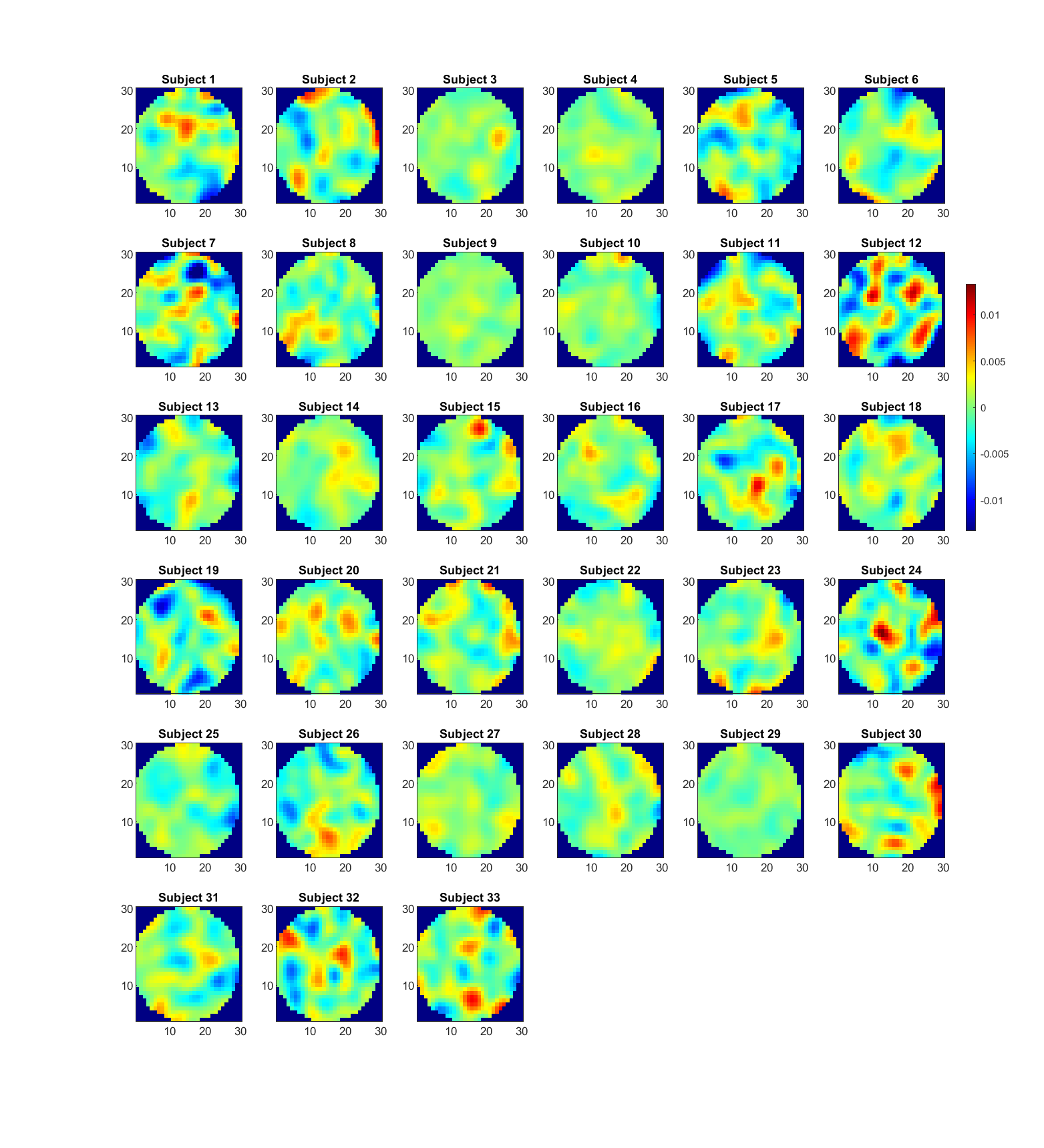}
    \caption{Subject-specific activation maps of the region of interest across the $33$ subjects.
    Note the differences in location of peaks across subjects, indicating significant inter-subject differences. }
    \label{fig:realData:raw}
\end{figure}

\begin{figure}
    \centering
    \includegraphics[scale = .55]{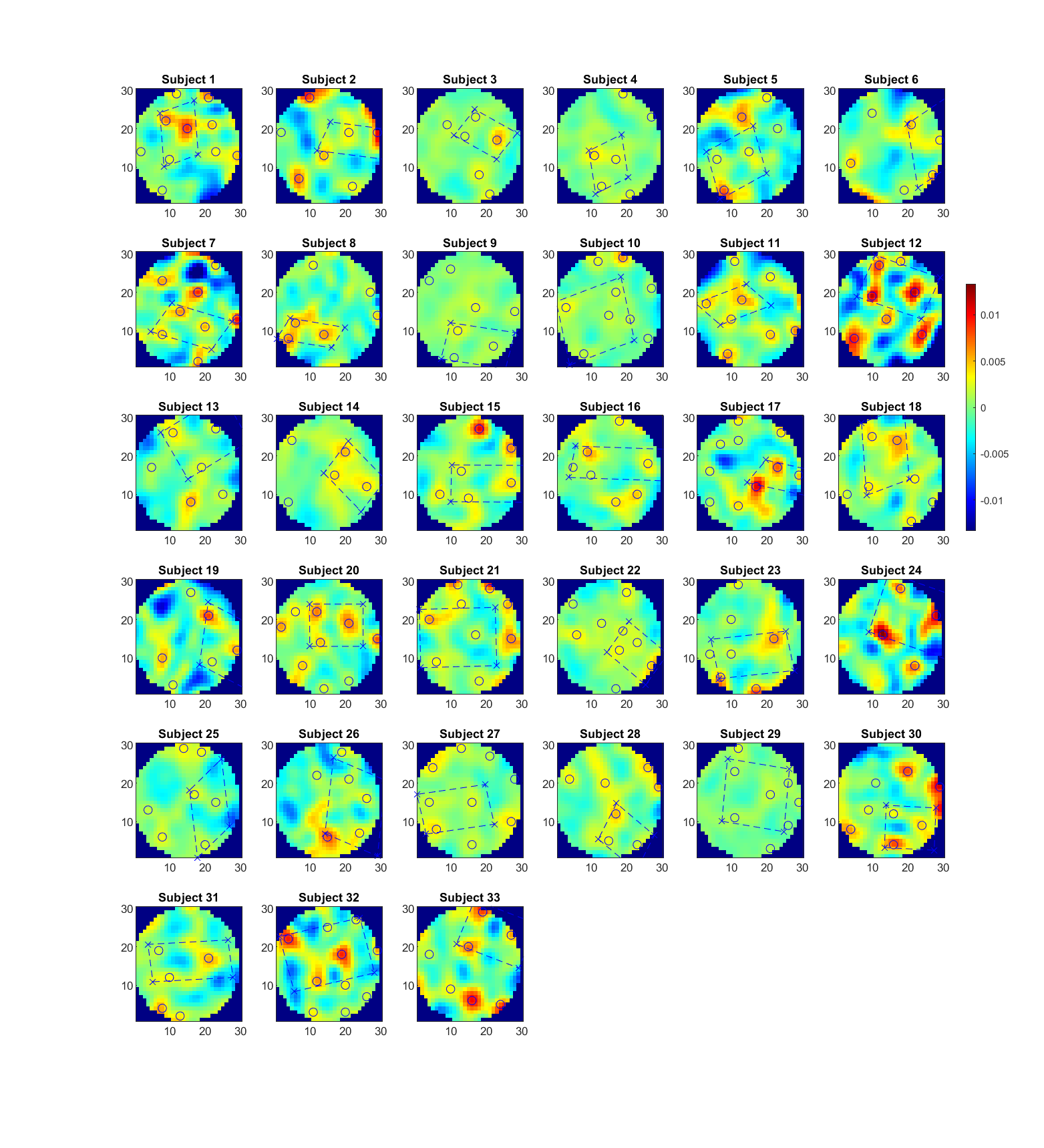}
    \caption{The corresponding cropped area (bounding box) based on prior information in each subject-specific map. The features (local peaks) are circled. }
    \label{fig:realData:prior}
\end{figure}

\begin{figure}
    \centering
    \includegraphics[scale = .55]{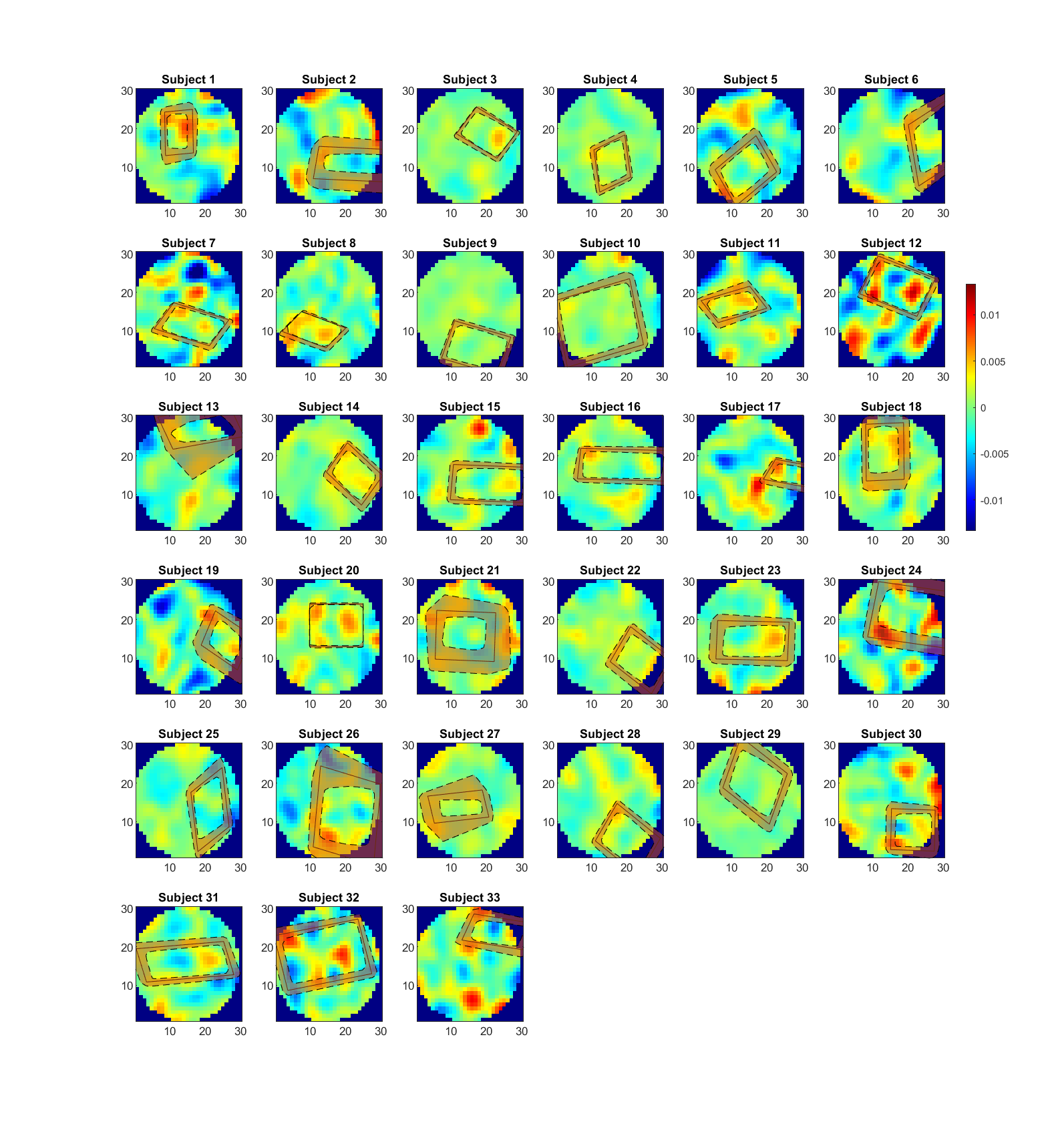}
    \caption{The warped bounding box using the transformation computed with the posterior mean is shown in solid red. The 95\% credible interval of the transformation map is illustrated in the shaded area. }
    \label{fig:realData:posterior}
\end{figure}


\begin{figure}
    \centering
    \includegraphics[scale=.55]{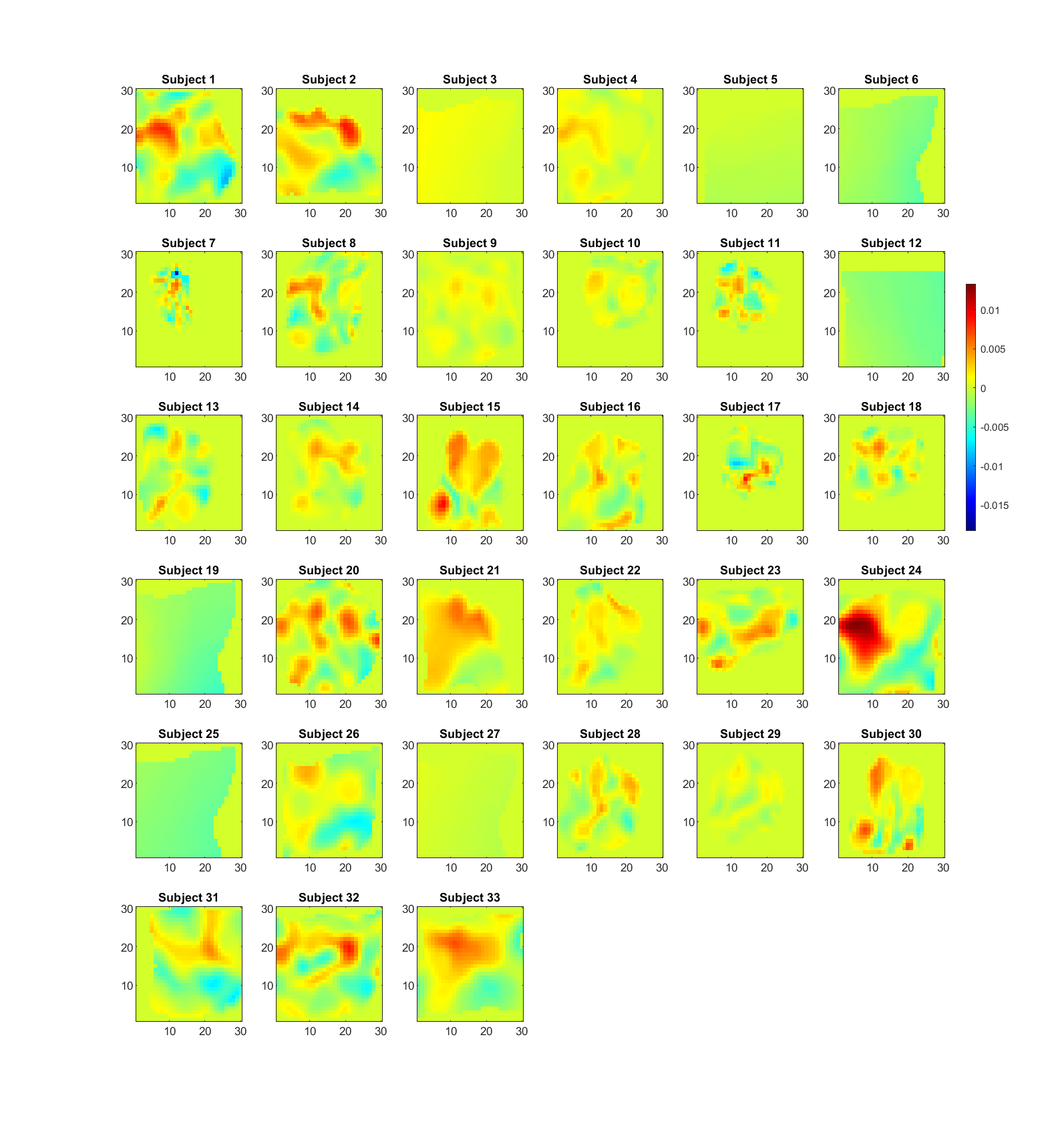}
    \caption{The warped images obtained using the diffeomorphic demons registration algorithm implemented in MATLAB's image processing toolbox. Note that multiple maps were not registered successfully, including subjects 3, 5, 6, 12, 19, 25, and 27.}
    \label{fig:demonsResults}
\end{figure}

\begin{figure}
    \centering
    \includegraphics[scale = .55]{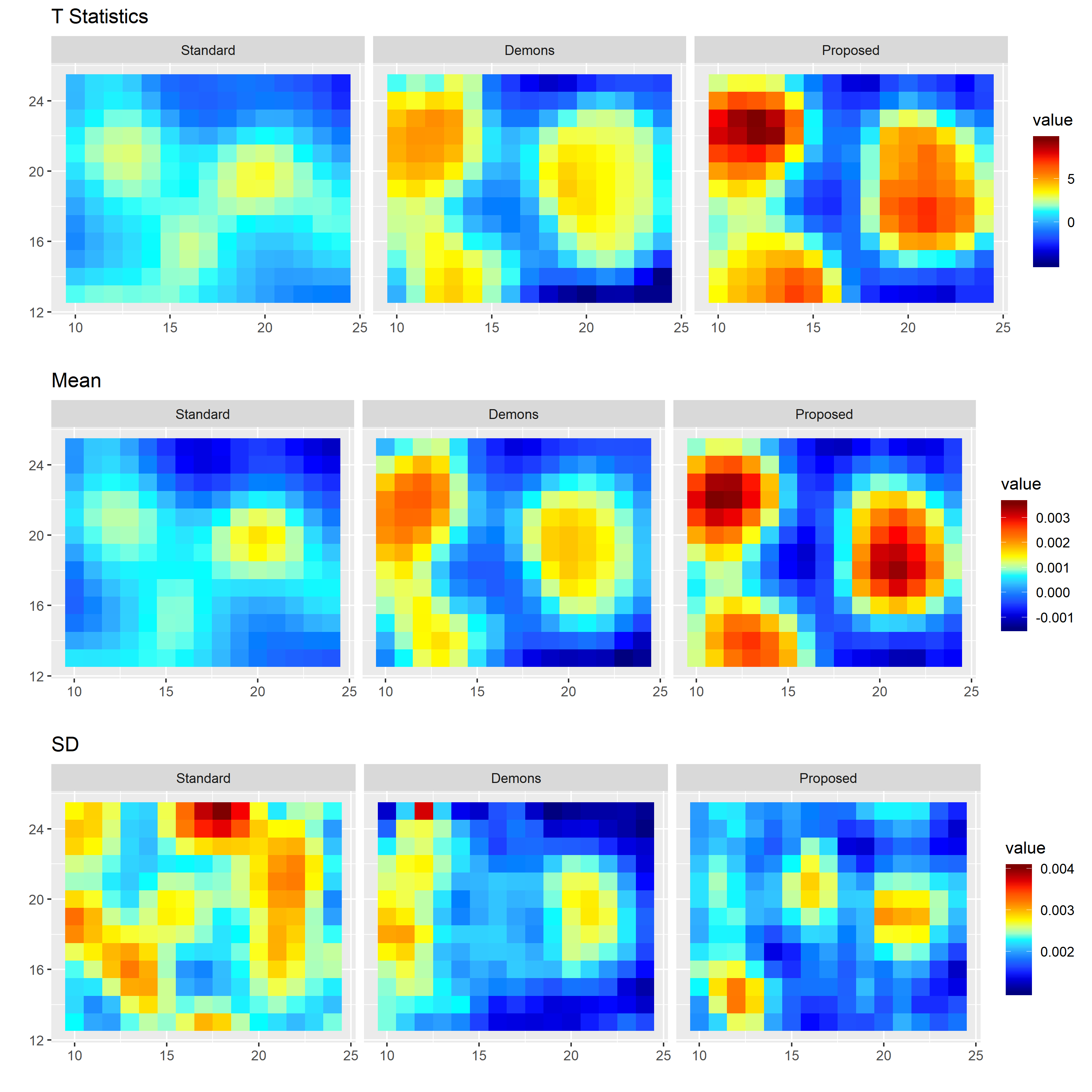}
    \caption{A comparison between the standard approach (left column), the diffeomorphic demons registration algorithm (middle column), and our proposed approach (right column).  Group-level statistics (t-statistics, group means, group standard deviations) are calculated for all three approaches. Our proposed approach shows more focal results and higher t-statistic values, indicating increased sensitivity for group-level inference.
    }
    \label{fig:realData:group analysis}
\end{figure}

\clearpage



%% file: supplement.tex
\beginsupplement

\begin{figure}
    \centering
    \includegraphics[scale = .7]{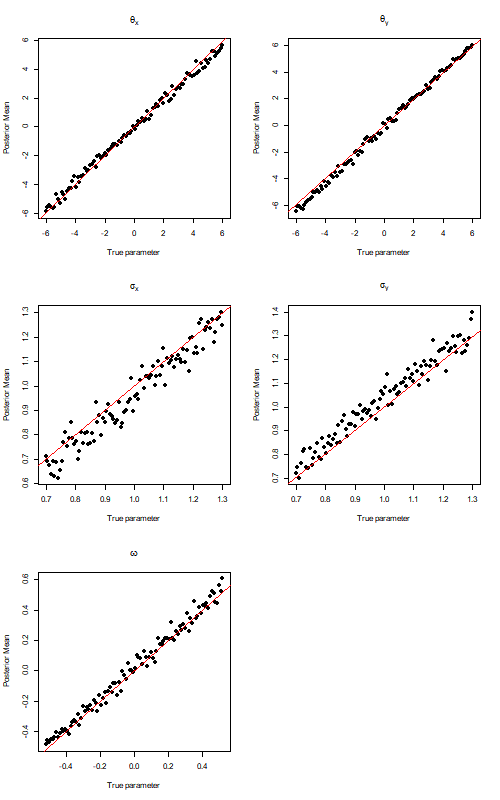}
    \caption{The posterior means plotted against the true parameters for each transformation parameter: translations (first row), scalings (second row), and rotations (third row).}
    \label{fig:nonlinear_simulation_100Comparison}
\end{figure}